\documentclass[journal=jpcafh, manuscript=article]{achemso}

\usepackage[version=3]{mhchem} % Formula subscripts using \ce{}
\usepackage{graphicx}% Include figure files
\usepackage{color}

\newcommand{\andy}[1]{ }

\newcommand{\beq}{\begin{equation}}
\newcommand{\eeq}{\end{equation}}
\newcommand{\barr}{\begin{eqnarray}}
\newcommand{\earr}{\end{eqnarray}}

\def\bra#1{\langle #1 |}
\def\ket#1{| #1 \rangle}
\def\cH{{\cal H}}
\def\bmp{\mbox{\boldmath $p$}}
\renewcommand{\Im}{\mathrm{Im}}
\author{D. Bruno}
\affiliation[CNR-IMIP]
{Istituto di Metodologie Inorganiche e dei Plasmi, 
Consiglio Nazionale delle Ricerche, Bari, Italy}
\author{P. Facchi}
\affiliation[Universit\`a di Bari-Matematica]
{Dipartimento di Matematica, Universit\`a di Bari, Bari, Italy}
\alsoaffiliation[INFN-Bari]
{Istituto Nazionale di Fisica Nucleare, Sezione di Bari, Bari, Italy}
\author{S. Longo}
\email{savino.longo@ba.imip.cnr.it}
\affiliation[Universit\`a di Bari-Chimica]
{Dipartimento di Chimica, Universit\`a di Bari, Bari, Italy}
\alsoaffiliation[CNR-IMIP]
{Istituto di Metodologie Inorganiche e dei Plasmi, 
Consiglio Nazionale delle Ricerche, Bari, Italy}
\author{P. Minelli}
\affiliation[CNR-IMIP]
{Istituto di Metodologie Inorganiche e dei Plasmi, 
Consiglio Nazionale delle Ricerche, Bari, Italy}
\author{S. Pascazio}
\affiliation[Universit\`a di Bari-Fisica]
{Dipartimento di Fisica, Universit\`a di Bari, Bari, Italy}
\alsoaffiliation[INFN-Bari]
{Istituto Nazionale di Fisica Nucleare, Sezione di Bari, Bari, Italy}
\author{A. Scardicchio}
\affiliation[ICTP]
{Abdus Salam International Centre for Theoretical Physics, Trieste, Italy}
\alsoaffiliation[INFN-Trieste]
{Istituto Nazionale di Fisica Nucleare, Sezione di Trieste, Trieste, Italy}

\title[QZE in a molecule]
{Quantum Zeno effect in a model multilevel molecule}

\begin{document}
%%%%%%%%%%%%%%%%%%%%%%%%%%%%%%%%%%%%%%%%%%%%%%%%%%%%%%%%%%%%%%%%%%%%%
%% The manuscript does not need to include \maketitle, which is
%% executed automatically.  The document should begin with an
%% abstract, if appropriate.  If one is given and should not be, the
%% contents will be gobbled.
%%%%%%%%%%%%%%%%%%%%%%%%%%%%%%%%%%%%%%%%%%%%%%%%%%%%%%%%%%%%%%%%%%%%%
\begin{abstract}
We study the dynamics of the populations of a model molecule
endowed with two sets of rotational levels of different parity,
whose ground levels are energy degenerate and coupled by a constant
interaction. The relaxation rate from one set of levels to the other
one has an interesting dependence on the average collision frequency of
the molecules in the gas. This is interpreted as a quantum Zeno
effect due to the decoherence effects provoked by the molecular
collisions.
\end{abstract}


\section{Introduction}
\label{sec-int}\andy{sec-int}
The quantum Zeno effect is usually formulated as the hindrance of
the evolution of a quantum system due to frequent measurements
performed by a classical apparatus \cite{Misra, Beskow} and is
formalized according to von Neumann projection rule \cite{von}. The
literature of the last few years on this topic is vast and
contemplates a variety of physical phenomena, ranging from
oscillating (few level) systems \cite{Cook} and alternative
proposals \cite{altriZeno} to \textit{bona fide} unstable systems
\cite{antiZeno}, where the so-called "inverse" Zeno effect can take place.

The ideas and concepts at the basis of the quantum Zeno effect (QZE)
were also successfully extended to continuous measurement processes
by different authors and in different contexts
\cite{Peres80} and led to a remarkable explanation of the
stability of chiral molecules
\cite{HS}. This was a fertile idea, in that it explained the
behavior of a variety of physical systems in terms of a similar
underlying mechanism.

The QZE is, however, a much more general phenomenon, that takes
place when a quantum system is strongly coupled to another system
\cite{PIO} or when it undergoes a rapid dephasing process. Such a
rapid loss of phase coherence ("decoherence") of the quantum
mechanical wave function (for instance as a result of frequent
interactions with the environment) is basically equivalent to a
continuous measurement process (the main difference being that the
state of the system is not necessarily explicitly recorded by a
pointer).

The quantum Zeno effect is always ultimately ascribable to the
short-time features of the dynamical evolution law \cite{review}: it
is only the study of this dynamical problem that determines the
range in which a frequent disturbance or interaction will yield a
QZE. The very definition of "frequent" is a delicate problem, that
depends on the features of the interaction Hamiltonian. Moreover,
one should also notice that the quantum system is not necessarily
frozen in its initial state \cite{compactregularize}, but rather
undergoes a "quantum Zeno dynamics", possibly evolving
\emph{away} from its initial state \cite{ZenoMP}. The study of such an evolution
in the "quantum Zeno subspace" \cite{theorem} is in itself an interesting problem, whose mathematical and physical aspects, 
as well as the possible applications
to chemistry and physical chemistry,
are not completely clear and require further study and elucidation.

After the seminal experiment by Itano and collaborators \cite{Cook},
the QZE has been experimentally verified in a variety of different situations, on
experiments involving photon polarization
\cite{kwiat}, nuclear spin isomers \cite{ChapovskyPRL}, individual ions
\cite{Balzer, Toschek, Wunderlich, balzer2002}, optical pumping
\cite{molhave2000}, NMR \cite{jones}, Bose-Einstein condensates
\cite{ketterle}, the photon number of the electromagnetic field in a cavity \cite{HR}, and new experiments are in preparation with neutron spin \cite{VESTA, RauchVESTA}.

We focus here on the interesting example of QZE proposed in
\cite{ChapovskyPRL}: the nuclear spin depolarization mechanisms in
$^{13}$CH$_3$F, due to magnetic dipole interactions and collisions
among the molecules in the gas, was experimentally investigated and
interpreted as a QZE. In a few words, the $^{13}$CH$_3$F molecule
has two kinds of angular momentum states, according to the value of
the total spin of the three protons (H nuclei): $I=3/2$ (ortho) and
$I=1/2$ (para). Transitions between states with different parity are
(electric dipole) forbidden, so that spin flip occurs via a weak
coupling between two levels of different spin parity (this is most
effective when there is an accidental degeneracy between the levels,
achievable, for example, via a Stark effect
\cite{Chapovskyrelax}). One observes a significant dependence of
the spin relaxation on the gas pressure and interprets this as a QZE
provoked by the dephasing due to molecular collisions. Nuclear spin
conversion in polyatomic molecules is reviewed in
\cite{Chapovskyreview}.

The aim of this article is to study the occurrence of the QZE in the
general framework of collision-inhibited Rabi-like oscillations
between two sets of rotational levels. We shall study the evolution
of the level populations in a model multilevel molecule endowed with two
sets of rotational levels of different parity. In particular, we
shall concentrate on the interesting effects that arise as a
consequence of the interactions (collisions) with the other
molecules in the gas. The model we shall adopt will be studied both
numerically and analytically, and the results will be compared. One
of the main objectives of our investigation will be the analysis of
apparently different phenomena in terms of a Zeno dynamics.

We shall introduce the system in Section \ref{sec-dpw} and the Zeno
problem in the present context in Section \ref{sec-zenos}. In
Sections \ref{sec-master}-\ref{sec-zenoeffcoupl} we study the
problem from an analytic point of view, by deriving and
approximately solving a master equation. In Section
\ref{sec-simulations} the analytical result are compared to an
accurate numerical simulation. We conclude in Section
\ref{sec-concrem} with a few remarks.

\section{The system}
\label{sec-dpw}
\andy{sec-dpw}
Our model molecule has two subsets of rotational levels (to be
called left ($L$) and right ($R$) levels in the following) of
different parity, whose ground levels are energy degenerate and
coupled by a constant interaction. (The choice of the ground levels
is motivated by simplicity: one could choose any other couple of
energy-degenerate levels in the $L$-$R$ subspaces.) The molecules
undergo collisions with other identical molecules in the gas and we
assume that these collisions couple the rotational energy levels but
do not violate spin parity conservation. We shall focus on the
dependence of the relaxation rate on the average collision time or,
equivalently, on gas pressure: a QZE takes place if the transition
between the left and right subspaces is inhibited when the
collisions become more frequent (i.e., the gas pressure increases).

\begin{figure}
%\begin{center}
\includegraphics{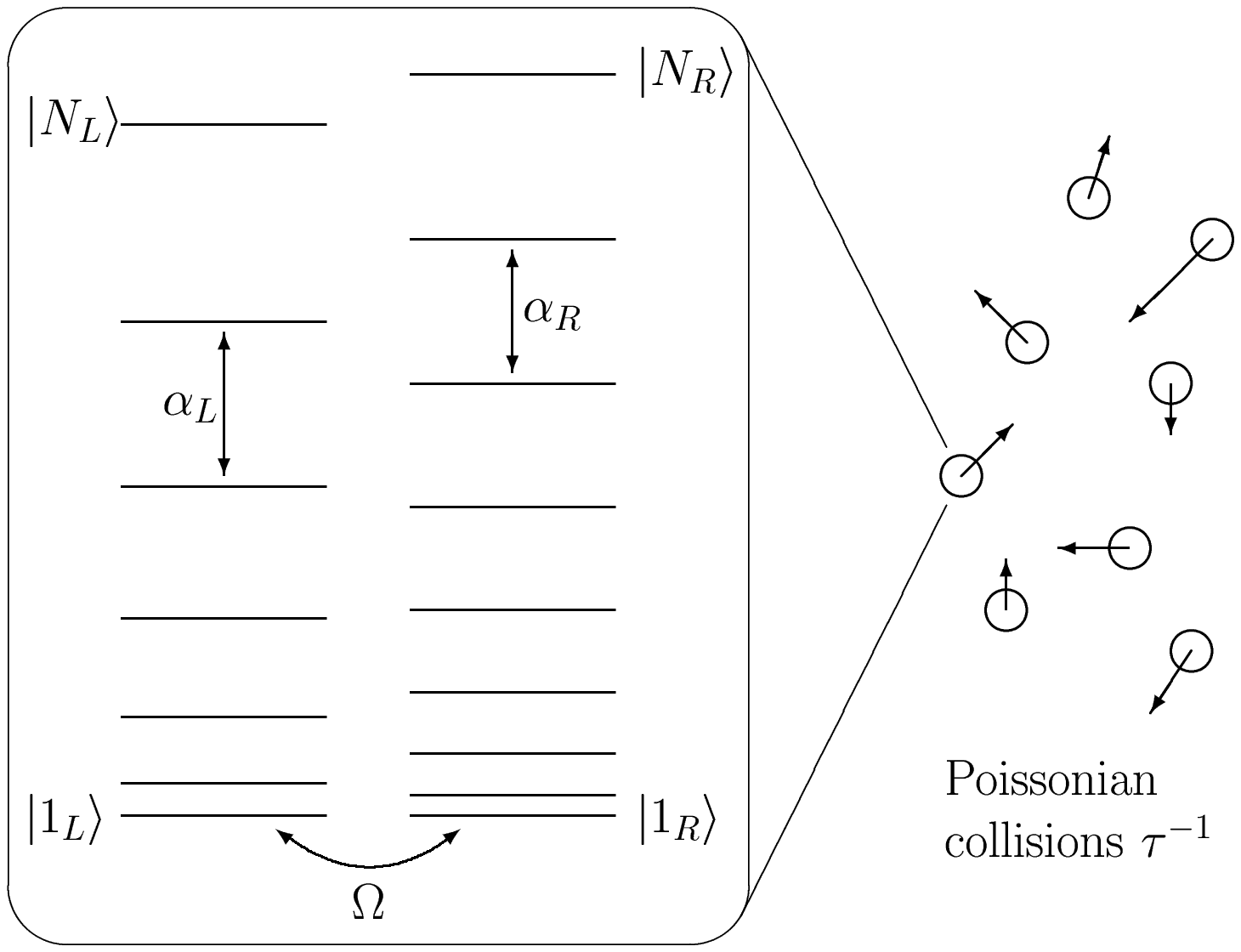}
%\end{center}
\caption{\label{fig:molecule}Poissonian collisions in a gas of multilevel molecules. }
%\label{fig:molecule}
\end{figure}

A sketch of the system is shown in \ref{fig:molecule}.
The total Hilbert spaces of each molecule is made up of two
subspaces $\cH_L$ (left) and $\cH_R$ (right),  with $N_L$ and $N_R$
levels respectively. Collisions cannot provoke $L
\leftrightarrow R$ transitions, so that no transitions are
possible between the two subspaces, except through their ground
states. However, collisions with other particles in the gas provoke
transitions \emph{within} each subspace.

The Hamiltonian is
\andy{Htot}
\beq
H= H_f+H_{\mathrm{coll}}(t)= H_0+H_1+H_{\mathrm{coll}}(t),
\label{eq:Htot}
\eeq
where $H_f=H_0+H_1$ is the free Hamiltonian and
\andy{Hdetails0,1,coll}
\barr
H_0 &=& \sum_{n_L=1}^{N_L} E_{n_L} \ket{n_L}\bra{n_L}
+\sum_{n_R=1}^{N_R} E_{n_R} \ket{n_R}\bra{n_R}, \
\label{Hdetail0} \\
H_1 &=& \hbar \Omega \left(\ket{1_L}\bra{1_R} +
\ket{1_R}\bra{1_L}\right),
\label{Hdetails1} \\
H_{\mathrm{coll}} &=& \hbar \sum_{j}\delta(t-\tau_j)\; V ,
\label{Hdetailscoll}\\
V &=&\alpha_L V_L+\alpha_R V_R,
\label{Hdetailscoll2}\\
V_s&=&\sum_{n_s=1}^{N_s-1} V_{n_s}
=\sum_{n_s=1}^{N_s-1}\ket{n_s}\bra{n_s+1}+\ket{n_s+1}\bra{n_s},
\;\;\;\;
\label{Hdetailscoll1}
\earr
with $s=L, R$. The energy levels $\ket{n_s}$ have energies $E_{n_s}$
($s=L, R$) and $H_1$ provokes $L \leftrightarrow R$ transitions
between the two ground states, with (Rabi) frequency $\Omega$.
$\Omega$ is small (in a sense to be made precise later), for such a
transition is electric-dipole forbidden. $H_{\mathrm{coll}}$
accounts for the effect of collisions with the gas (environment):
the collisions are distributed according to the Poisson statistics, appropriate for gas phase with short-range binary collisions,
so that they occur at times
\andy{avtau1}
\beq
\tau_{j+1}=\tau_j + \delta\tau_j,
\label{eq:avtau1}
\eeq
where $\delta\tau_j$'s are independent random variables with
distribution
\andy{avtau}
\beq
p(\delta\tau_j)=\frac{1}{\tau}\exp(-\delta\tau_j/\tau)
\label{eq:avtau}
\eeq
and (common) average $\tau$. The coupling constants $\alpha_{L, R}$
are in general different from each other and measure the
"effectiveness" of a collision. For the sake of simplicity we
assume that collisions provoke transitions only between adjacent
levels [$V_s$ in (\ref{Hdetailscoll1}) involves only "nearest
neighbors" couplings]. We will assume, for concreteness, that the
energy levels are rotational, so that
\andy{energyrot}
\beq\label{eq:energyrot}
E_{n_s} = \hbar \omega_s n_s (n_s+1) \quad (s=L, R)
\eeq
and $\ket{1_L}$ and $\ket{1_R}$ are the only resonant pair of
states:
\andy{resonant}
\beq
\label{eq:resonant}
E_{1_L}=E_{1_R}, \qquad E_{m_L} \neq E_{n_R} \quad
\mbox{for}\quad m_L, n_R > 1.
\eeq
See \ref{fig:molecule}. The Hilbert spaces $\cH_L$ and
$\cH_R$ are  finite dimensional, with  dimensions $N_L$ and $N_R$, respectively. This is
because, in general, the number of accessible rotational levels is
limited to a few tens, since for sufficiently high energies
molecules tend to dissociate. This could be accounted for by
introducing two "absorbing" levels $\ket{N_{L+1}},
\ket{N_{R+1}}$ \cite{absorbing}. However, in our analysis, we will
explore a time region in which the introduction of absorbing levels
is not necessary (in other words, the times involved will not be
long enough to display "border effects").

\section{Zeno effect}
\label{sec-zenos}\andy{sec-zenos}

Before we start our theoretical and numerical analysis it is
convenient to focus on the physics of the model introduced in the
preceding section and to clarify in which sense we expect a Zeno
effect to take place. We start from a simple numerical experiment
and 
 calculate the time evolution of the populations by the Monte Carlo method described in\cite{Savino1}.

Consider a uniform gas of identical molecules, having the internal
structure described in the preceding section. A single molecule
freely wanders in a total volume and undergoes random collisions. By
neglecting the spatial component of the wave function, each molecule
can be represented by an ($N_L+N_R$)-dimensional state vector
$\ket{\psi(t)}$ that describes its internal state
\cite{Savino3}. This physical situation is well schematized by the
model described in Sec. \ref{sec-dpw}. During the free flight the
evolution is governed by the free hamiltonian
\beq
\ket{\psi(t)}=\exp\left(-\frac{i}{\hbar} H_f t\right)\ket{\psi(0)}
.
\eeq
Since the molecules are immersed in a bath, the collisions are
distributed in time according to the Poisson statistics
(\ref{eq:avtau1})-(\ref{eq:avtau}), with average collision frequency
(per particle) $\tau^{-1}$. Once a collision occurs, a
collision time is sampled according to:
\beq
\delta\tau=-\tau \log(y),
\eeq
$y$ being a random number uniformly distributed in $[0, 1[$. The
collisions are modelled as instantaneous events and act on the
left/right subspaces independently. As a result of a collision, the
state becomes
\beq
\ket{\psi(t+0^+)}=\exp\left(-i\sum_{s=L, R}\alpha_s
V_s\right)\ket{\psi(t)} .
\eeq
The matrix $\exp(-i\sum_s\alpha_s V_s)$ is evaluated numerically. It
is assumed to be independent of the internal state of the colliding
partners and of their kinetic energy.

We also stress that since our aim is to investigate the occurrence
of a QZE \emph{within} the proposed level structure, we are not
interested in the dissociation of highly excited molecules. To this
end, we must restrict our attention to times such that the molecules
do not "see" the upper limit of the rotational levels, so that
"border" effects do not play any significant role. In this way the
dissociation of highly excited molecules can be safely neglected.

The afore-mentioned qualitative features of our analysis will be
carefully scrutinized and made precise in the following sections. We
now take them for granted and give a few preliminary results in
order to get a feeling for the physics at the basis of the Zeno
effect.

We set $N_L=N_R=40$ energy levels, with energies given by
(\ref{eq:energyrot}), where $ n_s=1, \dots, 40$, $\omega_L=1.3
\cdot 10^{10}\ {\mathrm{s}}^{-1}$ and $\omega_R=9.7 \cdot 10^9
{\mathrm{s}}^{-1}$. We always compute the average over an ensemble
of $5 \cdot 10^3$ particles. All particles are initially in the
$\ket{1_L}$ state and we study the temporal behavior of the relative
population in the left subspace
\andy{relleft}
\beq\label{eq:relleft}
P_L\equiv\sum_{n_L} p_{n_L},
\eeq
$p_{n_L}$ being the occupation probability of state $\ket{n_L}$.

\begin{figure}
\includegraphics{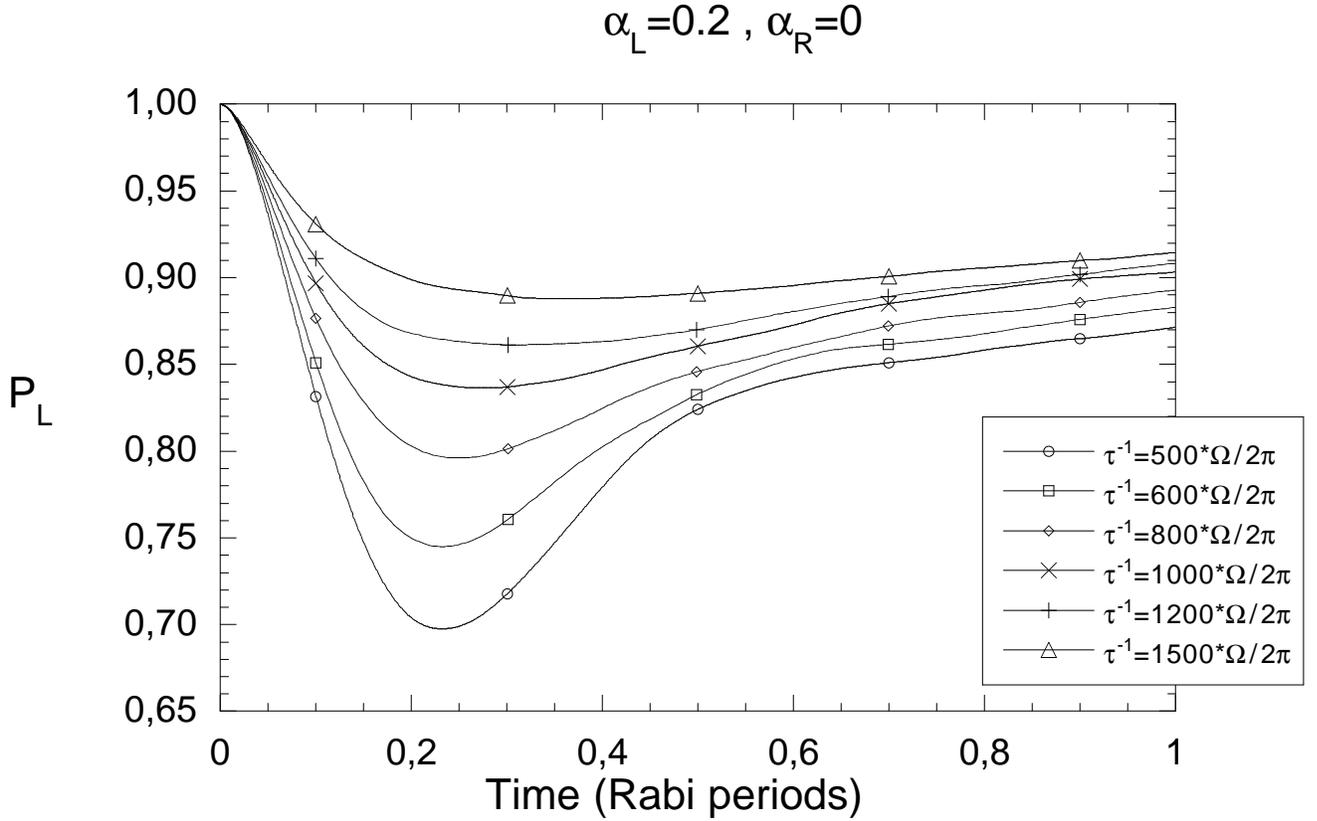}
\caption{\label{fig:zenoclass} Temporal evolution of $P_L$. The
collision frequency $\tau^{-1}$ is varied between $500T_R^{-1}$ and
$1500T_R^{-1}$ ($T_R=2\pi/\Omega$). We set $\alpha_L=0.2,
\alpha_R=0$, so that, in practice, $N_L=40$ left energy levels are
coupled to only $N_R=1$ right level. The survival probability in the
left subspace increases as the collision frequency is increased:
frequent collisions hinder transition to the right subspace, a
manifestation of a ("classically intuitive") Zeno effect. }
\end{figure}

\begin{figure}
\includegraphics{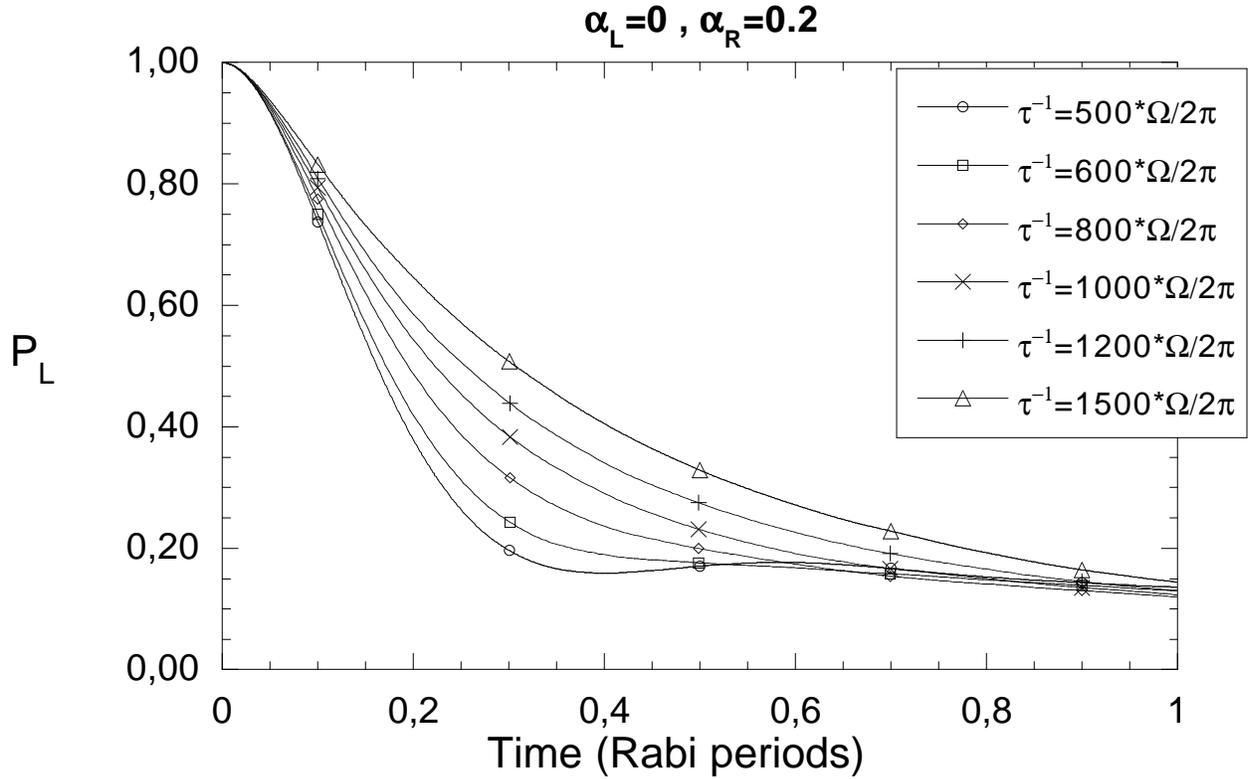}
\caption{\label{fig:zenoquant} Temporal evolution of $P_L$. The
collision frequency $\tau^{-1}$ is varied between $500T_R^{-1}$ and
$1500T_R^{-1}$. Unlike in the previous figure, we set $\alpha_L=0,
\alpha_R=0.2$, so that in practice, $N_L=1$ left level is coupled to
$N_R=40$ right levels. Again, the survival probability in the left
subspace increases as the collision frequency is increased: frequent
collisions hinder transition to the right subspace, a manifestation
of a ("classically counterintuitive") Zeno effect. }
\end{figure}

The results of our numerical integration are shown in 
\ref{fig:zenoclass}-\ref{fig:zenoquant}.
In \ref{fig:zenoclass}, $\alpha_L=0.2$ and $\alpha_R=0$, so
that collisions do not provoke transitions among the right states
(or, equivalently, the right subspace consists only of state
$\ket{1_R}$). It is apparent that when the collision frequency
$\tau^{-1}$ is increased between $500T_R^{-1}$ and $1500T_R^{-1}$
($T_R=2\pi/\Omega$ being the Rabi period) the survival probability
in the left subspace increases. If the collisions are viewed as a
dephasing process (effectively yielding a "measurement" of the
occupation probabilities of the left states), this can be viewed as
a Zeno effect. This is in agreement with our "classical" intuition:
since the system is initially in the left subspace and collisions
remove population density from the ground state $\ket{1_L}$ of this
subspace (the only level coupled to the right subspace), it is
intuitively clear that, by increasing the collision frequency,
transitions to the right subspace are hindered. This is a "classically intuitive" version of the Zeno effect.

The situation depicted in \ref{fig:zenoquant} is different:
here $\alpha_L=0$ and $\alpha_R=0.2$, so that now collisions do not
provoke transitions among the \textit{left} states, or equivalently
the left subspace consists only of state $\ket{1_L}$ (which is also
the only state coupled to the right subspace). Once again, when the
collision frequency $\tau^{-1}$ is increased in the same range as
before, the survival probability in the left subspace increases. We
stress that the collisions are effective in hindering the transition
from a single level towards a subspace that is \textit{initially
empty}. In other words, now the collisions act only on the right
subspace, where virtually no particles are present. Once again, this
can be viewed as a Zeno effect; however, it is somewhat less
intuitive than the previous one (and maybe a bit puzzling for our
"classical" intuition). This a "classically counterintuitive" version of the 
Zeno effect.

After having rapidly analyzed these two simple situations, we are
ready to tackle the more general case of $N_L$ left levels coupled
to $N_R$ right ones. This will be done in the following.

\section{Master equation}
\label{sec-master}\andy{sec-master}

\subsection{The general case}
\label{sec-mastergen}\andy{sec-mastergen}

We start our analysis by deriving a master equation for the density
matrix of the molecule. Write (\ref{Hdetailscoll}) as
\andy{Hcoll1}
\beq
H_{\mathrm{coll}}(t) = \hbar\; \mu(t) V,
\label{eq:Hcoll1}
\eeq
where
\beq
\mu(t)=\frac{d N(t)}{dt}=\sum_j \delta(t-\tau_j)
\eeq
is the derivative of a Poisson process $N(t)$ with mean time $\tau$
\cite{Gardiner}:
\beq
\mbox{Prob}\{N(t)=n\}=P(n, t)=e^{-t/\tau}
\frac{1}{n !}\left(\frac{t}{\tau}\right)^n .
\eeq
One gets
\andy{avprop}
\beq
\langle dN(t)\rangle=\frac{dt}{\tau}, \qquad
\left\langle\left(dN(t)-\frac{dt}{\tau}\right)^2\right\rangle=\frac{dt}{\tau},
\label{avprop}
\eeq
so that the process
\beq
dW(t)= \eta(t)dt= \mu(t) dt - \frac{dt}{\tau} =dN(t) -
\frac{dt}{\tau}
\label{aaa}
\eeq
has a vanishing mean and a linear variance in $dt$
\beq
\langle dW(t)\rangle=0, \qquad
\left\langle dW(t)^2\right\rangle=\frac{dt}{\tau}.
\eeq
In terms of the white noise $\eta(t)$ these equations read
\beq
\langle \eta(t)\rangle=0, \qquad
\left\langle \eta(t)\eta(t')\right\rangle=\frac{1}{\tau}\delta(t-t').
\eeq
The collision Hamiltonian can then be rewritten in terms of a
constant part and a white noise
\beq
H_{\mathrm{coll}}(t) = \frac{\hbar}{\tau} V + \hbar\; \eta(t) V,
\eeq
whence the total Hamiltonian (\ref{eq:Htot}) reads
\andy{newHtot}
\beq
H=\bar H + \hbar\; \eta(t) V, \qquad \bar
H=H_0+H_1+\frac{\hbar}{\tau} V.
\label{eq:newHtot}
\eeq
The Schr\"odinger equation (in It\^{o} form) is
\beq
\ket{\psi(t+dt)}=\left(1-\frac{i}{\hbar}\bar H dt -\frac{1}{2\tau}
V^2 dt \right)\ket{\psi(t)}-i  V dW \ket{\psi(t)}
\eeq
and the average density matrix [$\langle\cdots \rangle$ is
introduced in Eq.\ (\ref{avprop})],
\andy{eq:avdensm}
\beq
\rho(t)=\left\langle\;\ket{\psi(t)}\bra{\psi(t)}\;\right\rangle ,
\label{eq:avdensm}
\eeq
follows a master equation in the Kossakowski-Lindblad form \cite{Lindblad}
\barr
\frac{d\rho}{dt}&=&-\frac{i}{\hbar}\;[\bar H,
\rho]-\frac{1}{2\tau}\;[V,[V,\rho]]\nonumber\\
&=&-\frac{i}{\hbar}\;[\bar H,
\rho]-\frac{1}{2\tau}\;\{ V^2,\rho \} +
\frac{1}{\tau}\; V\rho V .
\earr
By using Eqs.\ (\ref{eq:newHtot}), (\ref{Hdetailscoll2}) and
(\ref{Hdetailscoll1}) we get
\andy{eq:genme}
\barr
\frac{d\rho}{dt}&=&-\frac{i}{\hbar}\;[ H_0, \rho]
-\frac{i}{\hbar}\;[H_1, \rho] -\frac{i}{\tau}\;[V, \rho]
-\frac{1}{2\tau}\;\{V^2,\rho\}+\frac{1}{\tau}\;V\rho V
\nonumber\\
&=& -\frac{i}{\hbar}\;[ H_0, \rho] -\frac{i}{\hbar}\;[H_1, \rho]
+\sum_{s=L, R}\left( -i\frac{\alpha_s}{\tau}[V_s,\rho] -
\frac{\alpha_s^2}{2\tau}\{V_s^2,\rho\}
+ \frac{\alpha_s^2}{\tau}
V_s\rho V_s \right) 
\nonumber\\
& & + \frac{\alpha_L \alpha_R}{\tau} \left(V_L
\rho V_R+V_R \rho V_L \right) .
\label{eq:genme}
\earr
This equation for the average density matrix (\ref{eq:avdensm}) is
exact, but complicated. However, it can be greatly simplified under
some reasonable hypotheses.

\subsection{Reduced master equation}
\label{sec-adiabatic}\andy{sec-adiabatic}

We assume that \textit{all} level pairs $E_{m_R}$ and $E_{n_L}$ are
sufficiently far from resonance, namely
\andy{reduced}
\beq
\frac{\Delta E}{\hbar} \gg \Omega,\ \tau^{-1},
\label{eq:reduced}
\eeq
where $\Delta E$ is the smallest energy difference between states
$\ket{m_R}$ and $\ket{n_L}$, with $m_R, n_L>1$. This requirement
will be discussed in more detail in Section \ref{sec-simulations}.
At this stage we only observe that, typically, $\Delta E/\hbar\simeq
10^{-9}$s, while $\Omega \simeq 1$kHz and $\tau
\simeq 1\mu$s, so the above condition appears very reasonable.

It is then possible to show that in (\ref{eq:genme}) the dynamics of
the populations $p_{m_s}= \rho_{m_sm_s}$ ($s=L, R$) plus the
coherence term $\rho_{1_L1_R}$ completely decouples from the
dynamics of the coherence terms $\rho_{m_s n_{s'}}$ ($s, s'=L, R$ and
$m_s, n_{s'}\neq 1$). This is because, roughly speaking, no
"diagonal" fast frequency is present [essentially because
$\bra{m_s}[H_0,\rho]\ket{m_s}=0$ in (\ref{eq:genme})] and, under
hypothesis (\ref{eq:reduced}), the contribution of all the other
fast terms is averaged to zero over the long timescales $\tau$ and
$\Omega^{-1}$, and the dynamics of the slow and fast terms
completely decouples. In conclusion, only the "slow" dynamics is
relevant over the large timescales $\tau$ and $\Omega^{-1}$.

The above argument has a general rigorous justification
\cite{theorem} in terms of an adiabatic theorem and is elucidated
in Appendix \ref{sec-appA} for the model studied in this article.
One shows that the part of the master equation (\ref{eq:genme})
pertaining to the populations becomes
\andy{eq:reducedme}
\beq
\label{eq:reducedme}
\frac{d \tilde\rho}{dt}\simeq
-\frac{i}{\hbar}\;[\tilde H_1, \tilde\rho] -\frac{i}{\tau}\;[\tilde
V,
\tilde\rho] -\frac{1}{2\tau}\;\{ \widetilde{V^2},\tilde\rho \} +
\frac{1}{\tau}\; \widetilde{ V  \tilde\rho  V} ,
\eeq
where the reduced operator $\tilde A$, defined by
\beq
\label{eq:Atilde}
\tilde A = Q A Q + \sum_{s=L, R}\sum_{m_s=2}^{N_s} P_{m_s} A
P_{m_s},
\eeq
involves only matrix elements belonging to the eigenspaces of $H_0$
\cite{theorem},
\barr
Q=P_{1_L}+P_{1_R} \qquad \mbox{and} \quad P_{m_s}=\ket{m_s}\bra{m_s}
,
\earr
[remember condition (\ref{eq:resonant})] and is diagonal with
respect to $H_0$
\beq
[H_0, \tilde A ]=0.
\eeq
In particular, from Eq.\ (\ref{Hdetails1})
\andy{eq:cohcontr}
\barr
\label{eq:cohcontr}
\tilde H_1  &=& Q H_1 Q = H_1 = \hbar \Omega \sigma_1, \nonumber\\
\sigma_1 &\equiv& \ket{1_L}\bra{1_R}+\ket{1_R}\bra{1_L}
\earr
and from Eqs.\ (\ref{Hdetailscoll2}) and (\ref{Hdetailscoll1})
\beq
\tilde V =0, \qquad \widetilde{V^2}=\sum_{s=L, R} \alpha_s^2
\widetilde{V_s^2}= \sum_{s=L, R} \alpha_s^2 \sum_{m_s=1}^{N_s-1}
V_{m_s}^2
\eeq
and
\beq
\widetilde{V\tilde\rho V}=\sum_{s=L, R} \alpha_s^2 \widetilde{ V_s
\tilde \rho V_s}=\sum_{s=L, R} \alpha_s^2 \sum_{m_s=1}^{N_s-1}
V_{m_s} \tilde \rho V_{ms} ,
\eeq
so that Eq.\ (\ref{eq:reducedme}) reads
\andy{eq:master}
\beq
\label{eq:master}
\frac{d \tilde\rho}{dt}= -i\Omega [\sigma_1,
\tilde \rho] -\sum_{s=L, R} \frac{\alpha_s^2}{2 \tau}
\sum_{m_s=1}^{N_s-1} [V_{m_s},[V_{m_s},\tilde\rho]].
\eeq
This is the master equation we will study in detail. The only
assumption made in its derivation is (\ref{eq:reduced}).

The reduced density matrix $\tilde \rho$ is given by Eq.\
(\ref{eq:Atilde}) and involves only the level populations
$p_{n_s}=\rho_{n_s n_s}$ ($s=L, R$) and the two coherence terms
$\rho_{1_L, 1_R}$ and $\rho_{1_R, 1_L}$, all other matrix elements
being zero. Thus, it describes two \emph{classical} Markov chains
$(1_L, \ldots, N_L)$ and $(1_R, \ldots, N_R)$, whose transition rates
are proportional to $D_L=\alpha_L^2/\tau$ and $D_R=\alpha_R^2/\tau$
respectively, linked by \emph{quantum} Rabi oscillations between
$\ket{1_L}$ and $\ket{1_R}$, whose period is $T_R=2\pi/\Omega$.

By setting $D=(D_L+D_R)/2$, i.e.\
$\alpha^2=(\alpha^2_L+\alpha^2_R)/2$, the ratio between the two
timescales
\andy{x}
\beq
\label{eq:x}
x= D T_R=\frac{\alpha^2}{\tau} T_R = \frac{2\pi
\alpha^2}{\tau\Omega}
\eeq
is an important parameter, that describes different dynamical
regimes. Larger values of $x$ correspond to more frequent collisions
(within a Rabi period) and consequently to a more evident
manifestation of the QZE.

\section{Stochastic dynamics in decoupled subspaces}
\label{sec-subspace}\andy{sec-subspace}

Let us first study the subdynamics of each subspace $\cH_{L/R}$
separately. To this end, set  $\Omega=0$ in Eq.\ (\ref{eq:master}):
the time evolution is governed only by the collision dynamics, the
right and left subspaces decouple and their subdynamics can be
studied separately.

In terms of ($s=L, R$)
\beq
\label{eq:ps}
\bmp^s=\left(\rho_{1_s 1_s},\ldots,\rho_{n_s
n_s,\ldots}\right)= \left(p_{1_s},\ldots,p_{n_s},\ldots \right),
\eeq
Eq. (\ref{eq:master}) reduces to
\andy{newevol}
\beq
\label{eq:newevol}
\frac{d \bmp^s}{dt} =  D_s W^s \bmp^s,
\eeq
where $W^s$ is the stochastic matrix
\beq
W^s=\left( \begin{array}{ccccc}
 -1 &  1 & 0 & 0 & \dots \\
  1 & -2 & 1 & 0 & \dots \\
   0 &  1 & -2 & 1 & \dots \\
   0 &  0 & 1 & -2 & \dots \\
   \vdots & \vdots & \vdots & \vdots & \ddots   \\
 \end{array} \right) ,
 \qquad (s=L, R)\ \
\label{eq:Ws}
\eeq
and
\beq
\label{eq:diffcoeff}
D_s=\frac{\alpha^2_s}{\tau} .
\eeq
Note that $W^s$ is a real symmetric matrix with real eigenvalues and
a complete set of eigenvectors.

The resulting dynamics is diffusive. Indeed, Eq.\ (\ref{eq:newevol})
explicitly reads
\barr
\dot p_{1_s} &=& D_s \left(- p_{1_s} + p_{2_s}\right),
\qquad \qquad (\Omega =0)
\label{eq:diffeqn1}\\
\dot p_{n_s} &=& D_s \left(p_{n_s-1} - 2 p_{n_s} +
p_{n_s+1}\right),
\quad (n_s\ge2) \label{eq:diffeqn2}
\earr
which is nothing but a diffusion equation (dropping the suffix $s$)
\beq
\label{eq:diffusion}
\partial_t p_n (t) = D\, \triangle p_n(t),
\eeq
where
\beq
\triangle \equiv \frac{1}{2}\left(\nabla^+\nabla^- +
\nabla^-\nabla^+\right) ,
\eeq
and
\beq
\nabla^+ p_n= p_{n+1}-p_n, \qquad \nabla^- p_n= p_n-p_{n-1}.
\eeq
The boundary condition $\nabla^- p_1 =0$ [see Eq.\
(\ref{eq:diffeqn1})] is imposed by introducing a supplementary state
$n=0$, whose probability satisfies $p_0=p_1$ for every $t$. The
evolution of the population is made up of two terms (both expressed
in terms of the fundamental solution of the heat equation): each
"site" (level) gets a direct and a "reflected" contribution from
the boundary $n=1$:
\andy{solutionp}
\beq
\label{eq:solutionp} p_n(t)=q_n(Dt) + q_{1-n}(Dt),
\eeq
where $q_n(t)$ are the probabilities of the continuous-time
symmetrical random walk engendered by the equations
\cite{VanKampen}
\andy{eq:diffeq}
\beq
\dot q_n = q_{n-1} - 2 q_n + q_{n+1}, \quad -\infty<n<\infty,
\label{eq:diffeq}
\eeq
whose solution starting at $n=1$ for $t=0$ [i.e.\ $p_n(0)=
\delta_{n 1}$] reads
\andy{solutionp1}
\beq
q_n(t)=e^{-2t} I_{|n-1|}(2t),
\label{eq:solutionp1}
\eeq
$I_n(t)$ being the modified Bessel function \cite{tables}. As is
well known, for $t\to\infty$ and $n\to\infty$ with $n^2/t=$const,
this yields
\beq
q_n(t)\sim\frac{1}{\sqrt{4 \pi t}}
\exp\left(-\frac{(n-1)^2}{4 t}\right).
\eeq
Note that the boundary condition is essential in assuring
probability conservation,
\beq
\sum_{n=1}^{+\infty}p_n(t)=\sum_{n=-\infty}^{+\infty} q_n(D t)=1 ,
\eeq
for any $t$. The above equations are of general validity. In
particular,
\andy{p1exact}
\beq
\label{p1exact}
p_1(t)= e^{-2 D t}
\left[I_0(2Dt)+I_1(2Dt)\right],
\eeq
(and $p_1(0)=1$). It is also possible, by using the solution
(\ref{eq:solutionp}) and (\ref{eq:solutionp1}), to evaluate the mean
and second moment
\barr
\label{def:mu}
\mu(t)&=&\sum_{n=1}^\infty  n p_n(t), \\
\label{def:sigma}
\sigma^2(t)&=&\sum_{n=1}^\infty n^2 p_n(t)\ .
\earr
Indeed, by using (\ref{eq:diffeq}), one can obtain explicit
differential equations involving these quantities, valid for any
$t$,
\barr
\dot{\mu}(t)&=& D p_1(t)= D e^{-2 D t}
\left[I_0(2Dt)+I_1(2Dt)\right],
\nonumber\\
\dot{\sigma^2}(t)&=&2 D + D p_1 (t),
\earr
whose integration gives
\barr
\mu(t)&=&\frac{1}{2}+\frac{1}{2}e^{-2 D
t}\left[(1+4Dt)\,I_0(2Dt)+4Dt\, I_1(2Dt) \right],
\nonumber\\ \label{eq:mut}\\
\label{eq:sigma2t}
\sigma^2(t)&=&2 D t + \mu(t).
\earr

Let us also give, for completeness, the expression of $\mu$ and
$\sigma$ for times $Dt\gg 1$ [but always $t\ll T_{\mathrm{d}}$, see
Eq.\ (\ref{eq:dissociation}) below]. From Eqs.\
(\ref{eq:mut})-(\ref{eq:sigma2t}),
\barr
\mu(t)&\sim&\sqrt{\frac{4 D}{\pi}t}, \nonumber\\
\sigma(t)&\sim&\sqrt{2 D t}.
\label{eq:mularge}
\earr In order to compare these
results with those of the following sections, consider that
\andy{Dt}
\beq
\label{eq:Dt}
Dt= x\frac{t}{T_R},
\eeq
where $T_R$ is the Rabi period and $x=x_s=\alpha_s^2 T_R/\tau \;
(s=1, 2)$ is essentially the scaling parameter introduced in
(\ref{eq:x}).

In reality, as we already emphasized, the number $N$ ($N_L$ or
$N_R$) of accessible rotational levels is in fact finite, because
the molecule dissociates after absorbing a sufficient amount of
energy. In order to account for this process one can add an
$(N+1)$th absorbing level (in each subspace). However, since we are
interested in phenomena, such as the QZE, that can be brought to
light within timescales shorter than the dissociation time, the
introduction of an absorbing level is an unnecessary complication
that can be easily avoided by restricting our attention to the
relevant timescales. Let us therefore estimate the time scale at
which dissociation occurs. If a molecule dissociates when it reaches
level $N+1$, namely if only $N$ levels take part in the diffusion
process, the "dissociation" time reads
\beq
\label{eq:dissociation}
T_{\mathrm{d}}\simeq \frac{N^2}{D}=\frac{N^2 \tau}{\alpha^2} =
\frac{N^2}{x} T_R.
\eeq
This is the time needed by the system, that starts in the ground
level, in order to reach the uppermost level via the diffusive
propagation engendered by the collision. This rough estimate of the
"dissociation" time can be improved: a better analysis yields
$T_{\mathrm{d}}=N^2 \tau/\pi^2 \alpha^2$, which is roughly of the
same order of magnitude.

In our analysis we will assume $N=40$. Within the numerical range of
the parameters $\alpha$ and $\tau$ to be used in our simulation, the
dissociation time $T_{\mathrm{d}}$ varies between $2$ and $12$ Rabi
periods. In the following we will always remain well below this
threshold.

\section{Zeno effect in coupled subspaces}
\label{sec-zenoeffcoupl}

We have seen in Section \ref{sec-zenos} that when the $L$ and $R$
subspaces are coupled, namely when $\Omega\neq 0$, a QZE can be
obtained by increasing the collision frequency. Indeed, as we will
show, by increasing the collision frequency, the probability of
remaining in the initial subspace decays more slowly. We also
commented on the possibility of studying the Zeno dynamics in two
different situations, one classically more intuitive and the other one less intuitive.
These different names reflect the fact that the former case can be
understood (at least qualitatively) by means of a classical Markov
process, while the latter cannot. Both Zeno effects are contained in
the master equation derived in Section \ref{sec-master} and are a
consequence of the features of the collisions with the other
molecules constituting the environment, or in other words, of the
coupling constants of the interaction Hamiltonian
$H_{\mathrm{coll}}$. The resulting dynamics will be numerically
investigated in full generality in Section  \ref{sec-simulations}.
However, before we show the results of the numerical simulation, let
us discuss the main qualitative features of the dynamics without
solving the complete master equation. This will be done in the
present section with the help of some working hypotheses and will
help us clarify some additional features of the Zeno effects.

When $\Omega\neq 0$ the two subspaces $\cH_L$ and $\cH_R$ are
coupled through their ground states. The evolution is described by
(\ref{eq:diffeqn2}), supplemented by the following three equations
\andy{diffeqn}
\barr
\label{eq:diffeqn}
\dot p_{1_L} &=& D_L \left(- p_{1_L} + p_{2_L}\right)+ \Omega
p^c(t) ,
\label{eq:diffeqn1L}\\
\dot p_{1_R} &=& D_R \left(- p_{1_R} + p_{2_R}\right)- \Omega
p^c(t) ,
\label{eq:diffeqn1R}\\
\dot{p}^c(t)&=&-\frac{D_L+D_R}{2}
p^c(t)-2\Omega\left(p_{1_L}(t)-p_{1_R}(t)\right), \ \
\label{eq:diffeqnpc}
\earr
where $p^c$ is the coherence term between states $\ket{1_L}$ and
$\ket{1_R}$
\beq
\label{eq:pceq}
p^c=-2 \Im \rho_{1_L 1_R}=i\left(\rho_{1_L 1_R}-\rho_{1_R
1_L}\right) ,
\eeq
responsible for the coupling between the two subspaces. The total
probabilities of being in the left and right subspaces read
\beq
P_L(t)=\sum_{n_L} p_{n_L}(t), \quad P_R(t)=\sum_{n_R} p_{n_R}(t),
\eeq
respectively. The derivatives of these quantities are easily seen to
be simply related to the coherence term:
\beq
\dot{P}_L(t)=\Omega p^c(t), \quad \dot{P}_R(t)=-\Omega p^c(t) .
\eeq
Notice that $\dot{P}_L+\dot{P}_R=0$ (conservation of particles
number). Let our particles start in the left subspace at time $t=0$.
Therefore the quantity of interest is $P_L$. One can obtain the
evolution equation for $P_L(t)$ by eliminating $p^c$ by means of
(\ref{eq:diffeqnpc})
\andy{eq:deqforPL}
\beq
\label{eq:deqforPL}
\ddot{P}_L+ D \dot{P_L}+2\Omega^2\left(p_{1_L}-p_{1_R}\right)=0,
\eeq
where we set $D\equiv (D_L+D_R)/2$. This equation shows that the
dynamics of $P_L$ is governed only by the population difference
between the ground states, irrespectively of the population of the
higher levels. This introduces an interesting picture of the
dynamics, in which the Rabi oscillations act as a "source" for the
probability. The source drains particles from the left to the right
subspace  if $p_{1_L}>p_{1_R}$ and \textit{vice versa} if
$p_{1_L}<p_{1_R}$.

Our initial condition will always be $p_{1_L}(0)=1$: initial
population in the ground state of the left subspace. For $t\ll
1/D\ll T_R=2\pi/\Omega$ (which is always true for our choice of
parameters), we can set $p_{1_L}(t)=1+O(Dt)$ and $p_{1_R}(t)=O(Dt)$
and a power-series solution of (\ref{eq:deqforPL}), with initial
conditions $P_L(0)=1,\
\dot{P}_L(0)=0$, yields
\andy{eq:ZenoT}
\beq
\label{eq:ZenoT}
P_L(t)=1-\Omega^2 t^2+o(\Omega^2 t^2),
\eeq
which shows that the quadratic region of the Rabi oscillation is not
perturbed by the collisional dynamics (namely, does not depend on
$\alpha$), even thought it extends up to times shorter than $1/D\ll
T_R$. This result was to be expected \cite{PIO} and is well observed
in our numerical experiments, also for very high collision
frequencies.

Equation (\ref{eq:deqforPL}) is exact, but it is not a closed
equation for the total probability $P_L$. One needs the populations
of the ground states in order to obtain $P_L$. We will therefore
introduce an ansatz for the functional form of the populations of
the ground states, valid for large values of the parameter $x$ defined in 
Eq.\ (\ref{eq:x}), which will enable us to get
a closed equation for $P_L$. In addition we will also gain a deeper
understanding of the Zeno phenomenon for this system. The ansatz
consists in substituting for $p_{1_{L/R}}$ the solution
(\ref{p1exact}), obtained for the decoupled subspaces ($\Omega=0$),
\emph{normalized} to $P_{L/R}$. This "adiabatic"
(Born-Oppenheimer-like) approximation relies upon the assumption
that the time scale of the internal collisional dynamics is much
faster than the Rabi one ($1/D\ll T_R$), so that particles are
drained from the ground level and redistributed according to the
uncoupled dynamics. The Rabi coupling simply accounts for the
varying number of particles present in each subspace. This ansatz is
translated into the equations
\andy{eq:ansatz}
\barr
p_{1_L}&=&P_L(t) f_L(t), \nonumber\\
p_{1_R}&=&P_R(t) f_R(t)=(1-P_L)f_R(t),
\label{eq:ansatz}
\earr
where $f_{L/R}(t)$ are the population probabilities of the ground
states given by the uncoupled dynamics (\ref{p1exact})
\andy{eq:findecsub}
\beq
\label{eq:findecsub}
f_{L/R}=e^{-2D_{L/R}t} \left[I_0(2D_{L/R}t)+I_1(2D_{L/R}t)\right].
\eeq
Substituting in (\ref{eq:deqforPL}) we obtain
\beq
\label{eq:pendulum}
\ddot{P}_L+D \dot{P}_L+2\Omega^2\left(P_L(f_L+f_R)-f_R\right)=0,
\eeq
which is the equation of motion of a unit-mass, forced pendulum with
varying frequency. The initial conditions are
$P_L(0)=1,\;\dot{P_L}(0)=0$. It is easy to prove that if
$f_R/(f_L+f_R)$ tends to a well-defined limit and its first and
second derivatives vanish when $t\to\infty$, there is a stable fixed
point at $t=\infty$ \footnote{Actually one always (implicitly)
assumes $t\ll T_{\mathrm{d}}$. If one looks at longer times $t
\geq T_{\mathrm{d}}$, this equilibrium state appears as a
metastable state, which decays into the true equilibrium state.}
\beq
P_L^*=\frac{f_R(t)}{f_R(t)+f_L(t)}\Bigg|_{t\to\infty}
\eeq
and any solution will eventually reach this point. This feature of
the population of the left subspace is always observed in the
numerical solutions. An asymptotic analysis of the Bessel functions,
performed with $1/D\ll t\ll T_{\mathrm{d}}$ shows that all these
requirements are satisfied and an equilibrium distribution exists
and is given by
\andy{eq:plstar}
\beq
\label{eq:plstar}
P_L^*=\frac{1}{1+\sqrt{D_R/D_L}} =
\frac{\alpha_L}{\alpha_L+\alpha_R}.
\eeq
Let us see now how the Zeno effect emerges in this picture in three
different cases.

\subsection{Case $\alpha_L=\alpha_R=\alpha\neq 0$}
\label{sec-zenoeffcouplA}

The first case-study is obtained by setting
$\alpha=\alpha_L=\alpha_R\neq 0$, so that the (collision dynamics in
the) two subspaces are identical and $f_L(t)=f_R(t)\equiv f(t)$. We
change the time variable from $t$ to the dimensionless $t/T_R=2\pi
t/\Omega$ and set $x=D T_R=\alpha^2 T_R/\tau=2\pi
\alpha^2\tau/\Omega$, obtaining (the dot denotes now
differentiation with respect to $t/T_R$)
\andy{eq:secordeqalfa}
\beq
\label{eq:secordeqalfa}
\ddot{P}_L+x\dot{P}_L+8\pi^2 f(t)(2P_L-1)=0,
\eeq
where $f(t)$ is given by (\ref{eq:findecsub}). Since $D_L=D_R$,
according to (\ref{eq:plstar}), $P_L$ will eventually tend to
$P_L^*=1/2$. However, we will see that the typical time scale
$T_{\mathrm{relax}}$ of this relaxation process will increase with
$x$ and this will be interpreted as a QZE.

The proposed analogy with a classical damped harmonic oscillator
suggests that when $x\gg 1$ we get $\dot{P}_L\sim 1/x^\alpha$ and
$\ddot{P}_L\sim 1/x^{2\alpha}$, with $\alpha>1$. Indeed, we will see
that the solution satisfies this hypothesis with $\alpha=3/2$, so
that the first term ($\ddot{P}_L\sim x^{-2\alpha}$) is negligible
with respect to the second ($x\dot{P}_L\sim x^{1-\alpha}$) and the
third one (both $f$ and $P_L$ are of order 1) and hence can be
dropped from (\ref{eq:secordeqalfa}). Thus we are left with a
first-order, separable differential equation whose solution is [here
$P_L(0)=1$ but $\dot{P}_L(0)=O(1/x)$]
\barr
P_L(t)&=&\frac{1}{2}+\frac{1}{2}\exp \left[
\frac{8\pi^2}{x^2}\Big(1-e^{-2xt}(1+4xt)\;I_0(2xt)
\right.\nonumber\\
& &\qquad\qquad\qquad\qquad \left. -e^{-2xt}4xt\;
I_1(2xt)\Big)\right].
\label{eq:equalPL}
\earr
For $xt\gg 1$ we obtain a stretched exponential
\beq
P_L(t)\simeq\frac{1}{2}+\frac{1}{2}\exp\left[\frac{8\pi^2}{x^2}-32
\pi^{3/2}\left(\frac{t}{x^3}\right)^{1/2}\right],
\label{eq:equalPLapp}
\eeq
from which one can define a relaxation time as the only
characteristic time present in the exponential (restoring natural
time units):
\beq
T_{\mathrm{relax}}\propto x^3 T_R .
\label{eq:equalzenoscale}
\eeq
The Zeno effect consists in the fact that by increasing $x$ (more
frequent collisions) the corresponding curves of $P_L$ tend to zero
more slowly. These predictions are in qualitative and quantitative
agreement with the numerical simulations of the next section.

\begin{figure}
\includegraphics{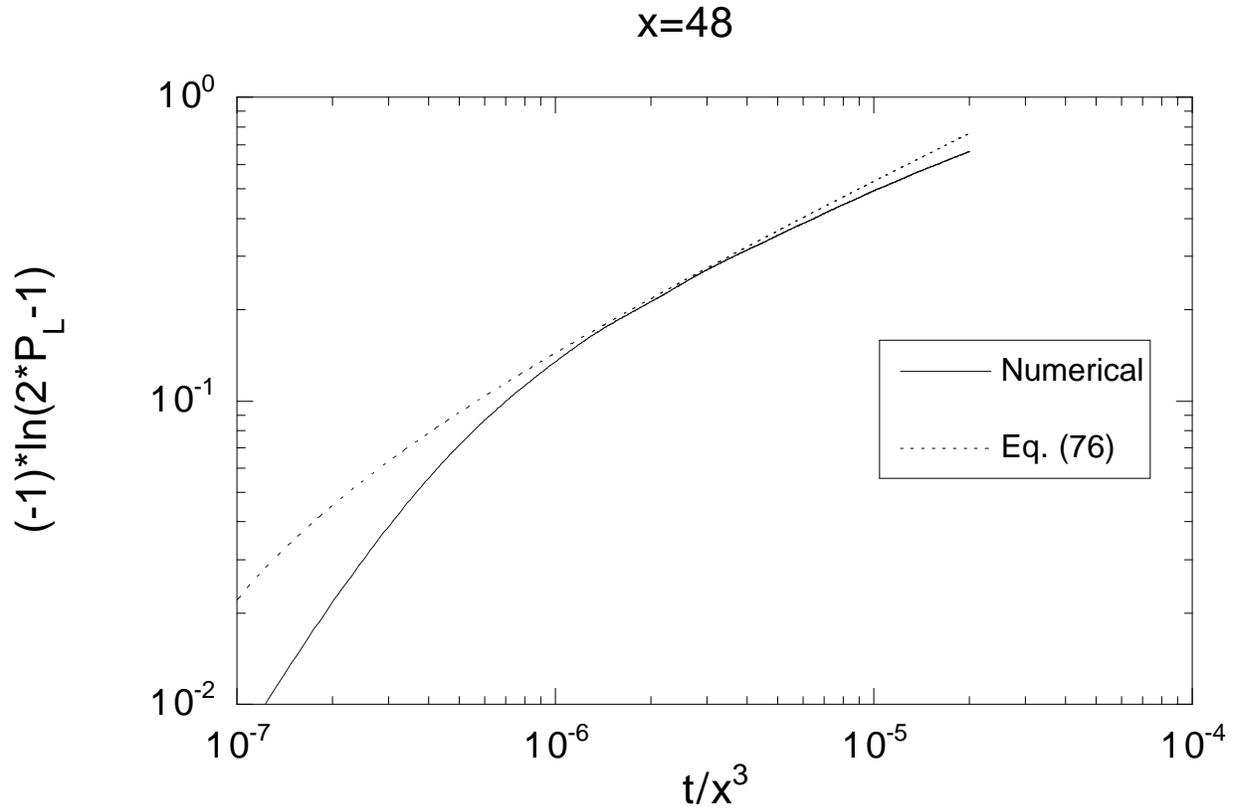}
\caption{\label{fig:fig77-3} Comparison between the numerical results and Eq.\
(\ref{eq:equalPLapp}). We set $x=48$.}
\end{figure}

\begin{figure}
\includegraphics{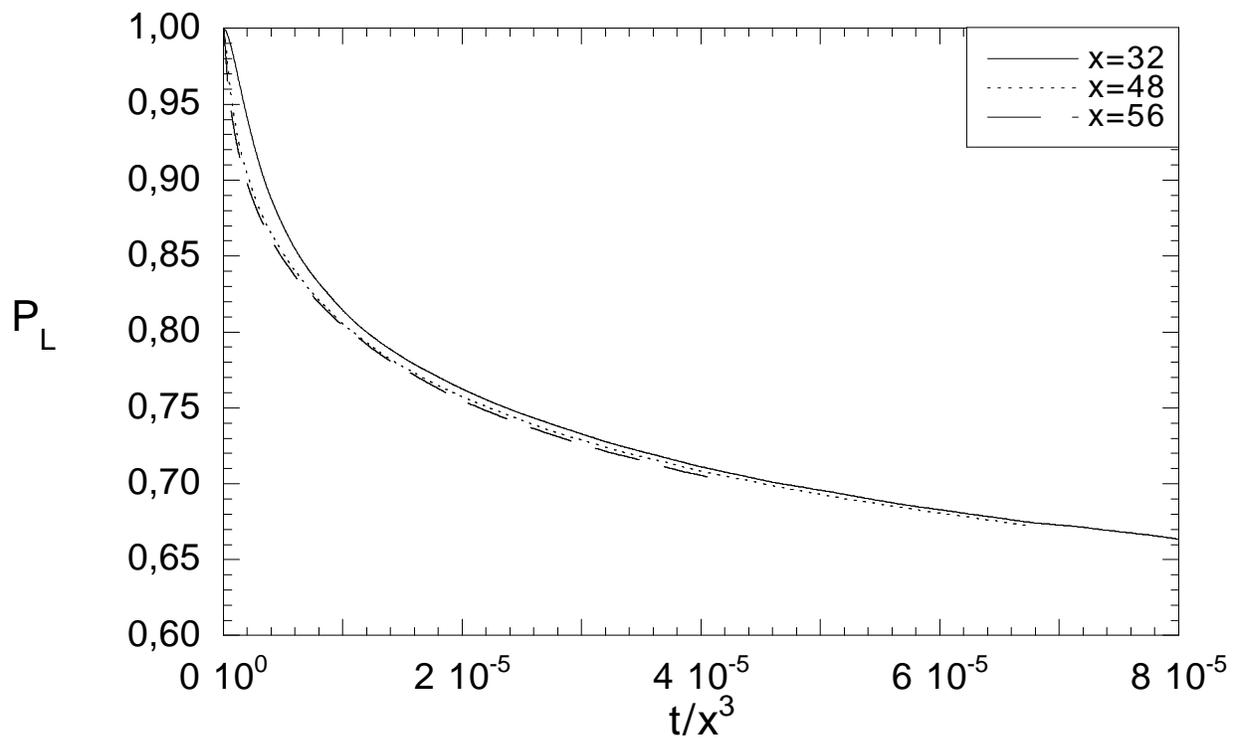}
\caption{\label{fig:sc} Rescaled probabilities for $x=32,\; 48,\;
56$ (numerical results).}
\end{figure}

In order to get a rough preliminary idea of the issues discussed in
this section, look for instance at
\ref{fig:fig77-3} and \ref{fig:sc}, where the numerical results
(to be described in greater details in the following) are compared
to Eqs.\ (\ref{eq:equalPLapp})-(\ref{eq:equalzenoscale}). The
probability (\ref{eq:equalPL})-(\ref{eq:equalPLapp}) is correct up
to a precision of $10\%$, showing that the ansatz (\ref{eq:ansatz})
yields sensible results. Notice that $x=48$ in
\ref{fig:fig77-3}, so that the solution (\ref{eq:equalPLapp}), which
is supposed to be valid for $xt
\gg1$, must yields accurate results for $t/x^3 \geq 10^{-6}$,
as one indeed observes. A numerical fit for the exponent in the
stretched-exponential yields $t^{0.3}$ rather than $t^{1/2}$,
confirming the general functional dependence. The very fact that the
global relaxation law is of the stretched-exponential type suggests
that the dynamics is highly nontrivial, but we will not elaborate on
this here. Finally, as can be seen from
\ref{fig:sc}, the scaling law (\ref{eq:equalzenoscale}) is very
well verified.

\subsection{Case $\alpha_L\neq \alpha_R = 0$}
\label{sec-zenoeffcouplB}

Let us briefly reconsider the first case analyzed in Sec.\
\ref{sec-zenos}, 
\ref{fig:zenoclass}. Here the left subspace is affected by
collisions while the right one is not. Although this is not a
realistic situation, it is interesting and instructive to look at
it. We shall show that also in this case, as the collision strength
is increased, the system tends to spend more time in the initial
(left) subspace.

If $\alpha_R=0$ and $\alpha_L\neq 0$ then $f_L\equiv f$ and $f_R=1$
and Eq.\ (\ref{eq:pendulum}) reads
\beq
\ddot{P}_L+x\dot{P}_L+8\pi^2[P_L(1+f(t))-1]=0,
\eeq
where $x=\alpha_L^2 T_R/2\tau=\pi\alpha_L^2/\Omega \tau$. By means
of the same approximations of the preceding subsection we obtain,
for $x\gg 2\sqrt{2}\pi$,
\barr
P_L(t)&\simeq& 1-\frac{2\sqrt{2}\pi}{x}e^{-\left(\sqrt{\frac{8\pi^2
t}{x}}+\frac{2\sqrt{2\pi}}{x}\right)^2} \nonumber\\
\label{eq:secondPLappsol}
&& \times \Phi\left(\sqrt{\frac{8\pi^2
t}{x}}+\frac{2\sqrt{2\pi}}{x}\right),
\earr
where $\Phi(z)$ is the error function of imaginary argument
\cite{tables}
\beq
\Phi(z)=\frac{2}{\sqrt{\pi}}\int_0^z dx e^{x^2}.
\eeq
Here the definition of a relaxation time is not easy (no simple
scaling law exists). However, both in this solution and in the
numerical data, $P_L$ has a single minimum $P^*_L$, which is an
increasing function of $x$: this can be regarded as a manifestation
of a (classically intuitive) Zeno effect, as explained in Sec.
\ref{sec-zenos}. From (\ref{eq:secondPLappsol}) the value of the
minimum is
\andy{eq:minimo}
\beq
\label{eq:minimo}
P_L^*=1-\frac{2.7}{x}
\eeq
and is an increasing function of $x$ \footnote{Consider
$f(y)=e^{-y^2}\Phi(y)$. Then the numerical value $2.7$ in
(\ref{eq:minimo}) is given by $2\sqrt{2}\pi f^*$, where $f^*=0.621$
is the maximum of $f$.}. This law is well confirmed by the numerical 
results shown
in \ref{fig:zenoclass}. Beyond the minimum $P_L$ tends
to 1 with a power-law
\beq
P_L(t)\simeq 1-\sqrt{\frac{4}{\pi x t}}\; .
\eeq
This is again a Zeno effect: by increasing the collision rate $x$
the survival probability increases.

\subsection{Case $\alpha_R\neq \alpha_L = 0$}
\label{sec-zenoeffcouplC}

This is the second case analyzed in Sec. \ref{sec-zenos},
\ref{fig:zenoquant}. If $\alpha_L=0,\ \alpha_R\neq 0$, then
$f_R=f, f_L=0$ and Eq.\ (\ref{eq:pendulum}) reads (here
$x=\alpha_R^2 T_R /2\tau$)
\beq
\ddot{P}_L+x\dot{P}_L+8\pi^2[P_L(1+f(t))-f(t)]=0.
\eeq
Again we neglect $\ddot{P}_L$ with respect to $x\dot{P}_L$ and
$P_L$, obtaining a first-order equation whose solution is (in the
large $x$ limit)
\barr
P_L(t)&\simeq& e^{-\left(\sqrt{\frac{8\pi^2
t}{x}}+\frac{2\sqrt{2\pi}}{x}\right)^2}\nonumber\\
&& \times
\left[1+\frac{2\sqrt{2}\pi}{x}\Phi\left(\sqrt{\frac{8\pi^2
t}{x}}+\frac{2\sqrt{2\pi}}{x}\right)\right].
\earr
This displays a (quantum) Zeno effect, since for $xt\gg 1$ one gets
\beq
P_L(t)\sim e^{-\left(\sqrt{\frac{8\pi^2
t}{x}}+\frac{2\sqrt{2\pi}}{x}\right)^2}
\eeq
[compare with (\ref{eq:secondPLappsol})].

Once again there is a scaling law and one can define a
characteristic relaxation time (in natural units)
\beq
T_{\mathrm{relax}}\sim x T_R .
\eeq
Observe that this scaling is at variance with
(\ref{eq:equalzenoscale}).

\section{Simulations}
\label{sec-simulations}\andy{sec-simulations}

\subsection{Method}
\label{sec-implement}\andy{sec-implement}
We will now study in detail the features and results of the
integration of the kinetic equation by means of a Monte Carlo  method
already used in the past to study the kinetics of two-level systems in nonequilibrium gases \cite{Savino1, Savino3}.

Let us recall the main features of the simulation. Some details have
already been given in Sec. \ref{sec-zenos}. We set $\Omega=935\
{\mathrm{s}}^{-1}$, $\alpha=\alpha_R=\alpha_L\simeq 0.2\div 0.4$,
$\tau^{-1} \leq 2 \cdot 10^5\ {\mathrm{s}}^{-1}$ and $N_L=N_R= 40$
energy levels in each subspace, with energies given by
(\ref{eq:energyrot}), where $ n_s=1,\dots, 40$, $\omega_L=1.3 \cdot
10^{10}\ {\mathrm{s}}^{-1},\ \omega_R=9.7
\cdot 10^9 {\mathrm{s}}^{-1}$. The minimum energy difference
$\Delta E$ between the levels is of great importance. One can check
that with the above-mentioned numerical figures $\Delta E/\hbar=2.8
\cdot 10^9 {\mathrm{s}}^{-1}$ and the condition (\ref{eq:reduced})
is always satisfied \footnote{The determination of $\delta
E\equiv\min_{1\leq m_L\leq N_L,\ 1\leq n_R\leq N_R}
|E_{m_L}-E_{n_R}|$ for generic $N_{L/R}$, with $E_{n_s}$ given by
(\ref{eq:energyrot}), poses an interesting problem of number theory.
However, in our case $N_L=N_R=40$ and one can numerically check that
the value $\Delta E/\hbar=2.8\cdot 10^9$s$^{-1}$ given in the text
is stable against perturbation of $\omega_{L, R}$ of a few percent
(well above experimental uncertainties).}. The populations dynamics
is collected as an average over an ensemble of $5 \cdot 10^3$
simulated particles. Since the underlying equations are linear, the
particles can be serially simulated and the precision of the results
sharpened by simply increasing the sample size. The simulations
provide the time variation of all the elements of the 1-particle
reduced density matrix. We constantly checked all the level
populations $p_{n_s}$, $1\leq n_s\leq 40$, $s=L, R$, but will only
discuss in the following the temporal behavior of the total
population of the left subspace $P_L$. The initial situation, in all
the simulations, is
\beq
p_{1_L}=1,\quad \text{all others} =0,
\eeq
so that the initial population is concentrated in the $\ket{1_L}$
state (the ground state of the left subspace).

\subsection{Results}
\begin{figure}
\includegraphics{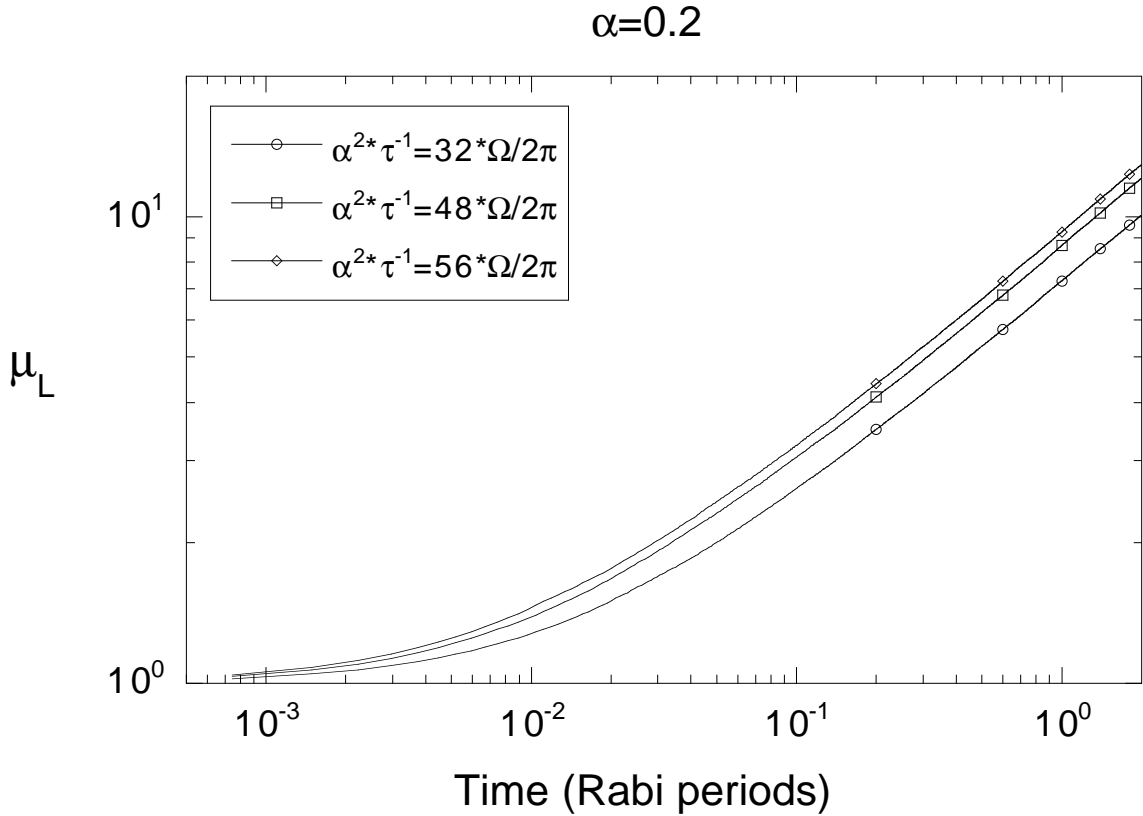}
\caption{\label{fig:fig5} Temporal evolution of the mean $\mu_L$
introduced in Eq.\ (\ref{def:mu}), for
$\alpha_R=\alpha_L=\alpha=0.2$ and $x=32,\ 48,\ 56$.}
\end{figure}

\begin{figure}
\includegraphics{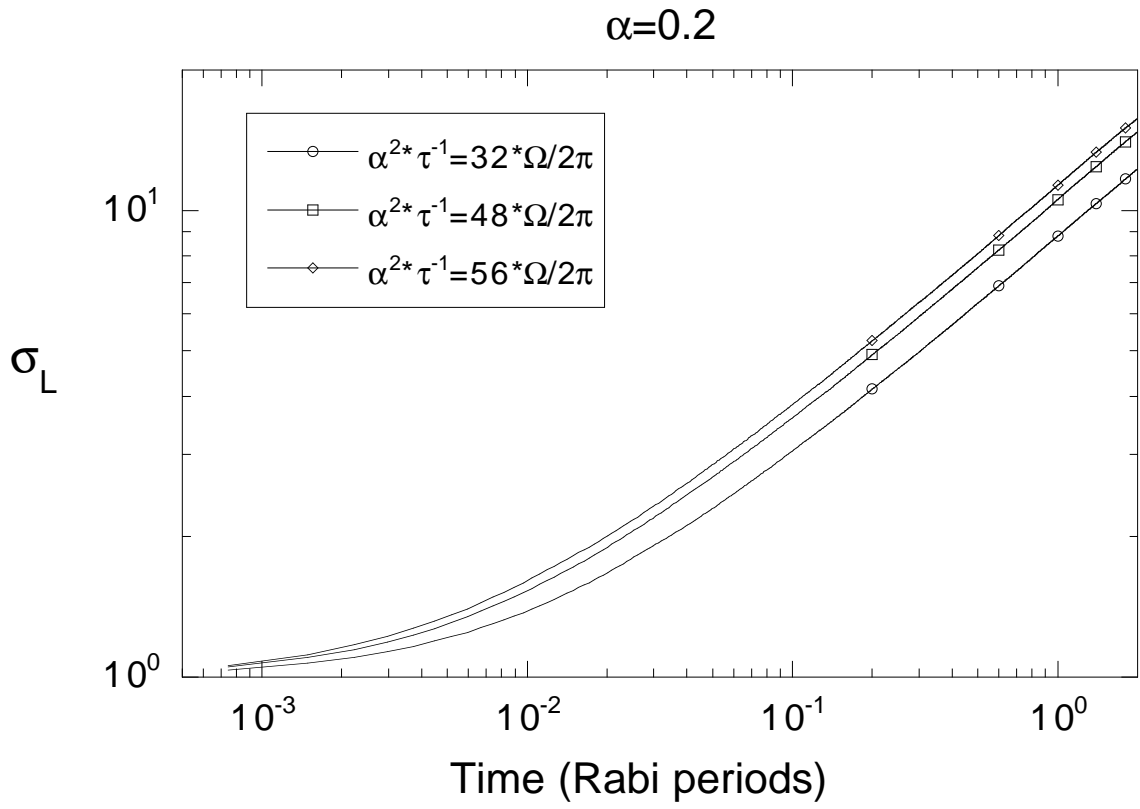}
\caption{\label{fig:fig6} Temporal evolution of the standard
deviation $\sigma_L$ introduced in Eq.\ (\ref{def:sigma}), for
$\alpha_R=\alpha_L=\alpha=0.2$ and $x=32,\ 48,\ 56$.}
\end{figure}

It is interesting to discuss in more detail some features of the
relaxation process and compare them to the analytical model proposed
in Sec.\ \ref{sec-master}. We track the temporal evolution of all
the populations and try to estimate the speed and the extent at
which the levels get populated. Two suitable indicators are the mean
$\mu=\mu_L$ and standard deviation $\sigma=\sigma_L$, introduced in
(\ref{def:mu}) and (\ref{def:sigma}). They are plotted in 
\ref{fig:fig5} and
\ref{fig:fig6} and accurately reproduce the analytical results
(\ref{eq:mut}) and (\ref{eq:sigma2t}) (remember that $Dt= x t/
T_R$). The analytical results are not shown in the graphs, for they
cannot be distinguished from the numerical ones.

Notice also in both figures the square-root dependence
(\ref{eq:mularge}) for large times $t\gg T_R/x\simeq 3 \cdot
10^{-2}T_R$. It is worth stressing that this also provides a direct
proof that boundary effects, related to the finiteness of the number
of levels, can be safely neglected for the times considered here.


\begin{figure}
\includegraphics{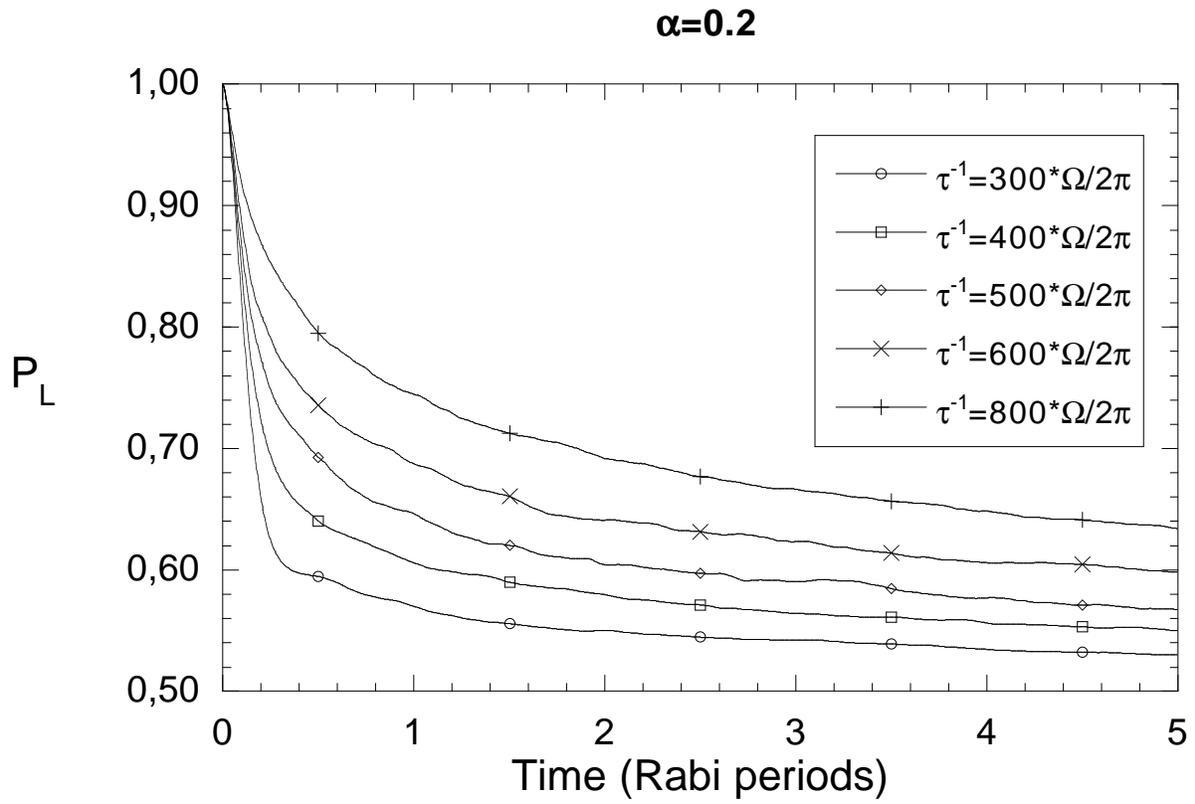}
\caption{\label{fig:fig1} Temporal evolution of $P_L$ as a
function of the collision frequency $\tau^{-1}$. We always set
$\alpha=\alpha_l=\alpha_R=0.2$. }
\end{figure}

We now show how the relaxation of the population depends on the
collision frequency $\tau^{-1}$, for fixed values of the parameter
$\alpha=\alpha_L=\alpha_R$.  \ref{fig:fig1} shows the temporal
evolution of the relative population of the left subspace
$P_L(t)\equiv\sum_{n_L} p_{n_L}(t)$ (once again, the analytical
results cannot be distinguished from the numerical ones and are not
shown in the graph). We note that this quantity will always
eventually tend to its equilibrium value $P_L^*=1/2$, according to
(\ref{eq:plstar}). However, the important point is that by
increasing the collision frequency from $300\ T_R^{-1}$ to $800\
T_R^{-1}$, the system tends to remain in the left subspace for a
longer time. This is evident in the plot and is a clear
manifestation of a QZE. We also notice (although this is not clearly
visible in 
\ref{fig:fig1}, due to the scale chosen) that there is always a
short-time quadratic region, characterized by a "Zeno time" $\ddot
P_L (0)=-\Omega$, in full agreement with Eq.\ (\ref{eq:ZenoT}). The
features of this short-time region are independent of other
parameters (such as $\alpha$ and $\tau$)
\cite{PIO}, as can be seen in the figure. Finally, we emphasize
that $x=\alpha^2 T/\tau$ ranges between 12 and 32 and is therefore
always $\gg 1$, so that the analysis of Sec.\
\ref{sec-zenoeffcouplA} applies.

\begin{figure}
\includegraphics{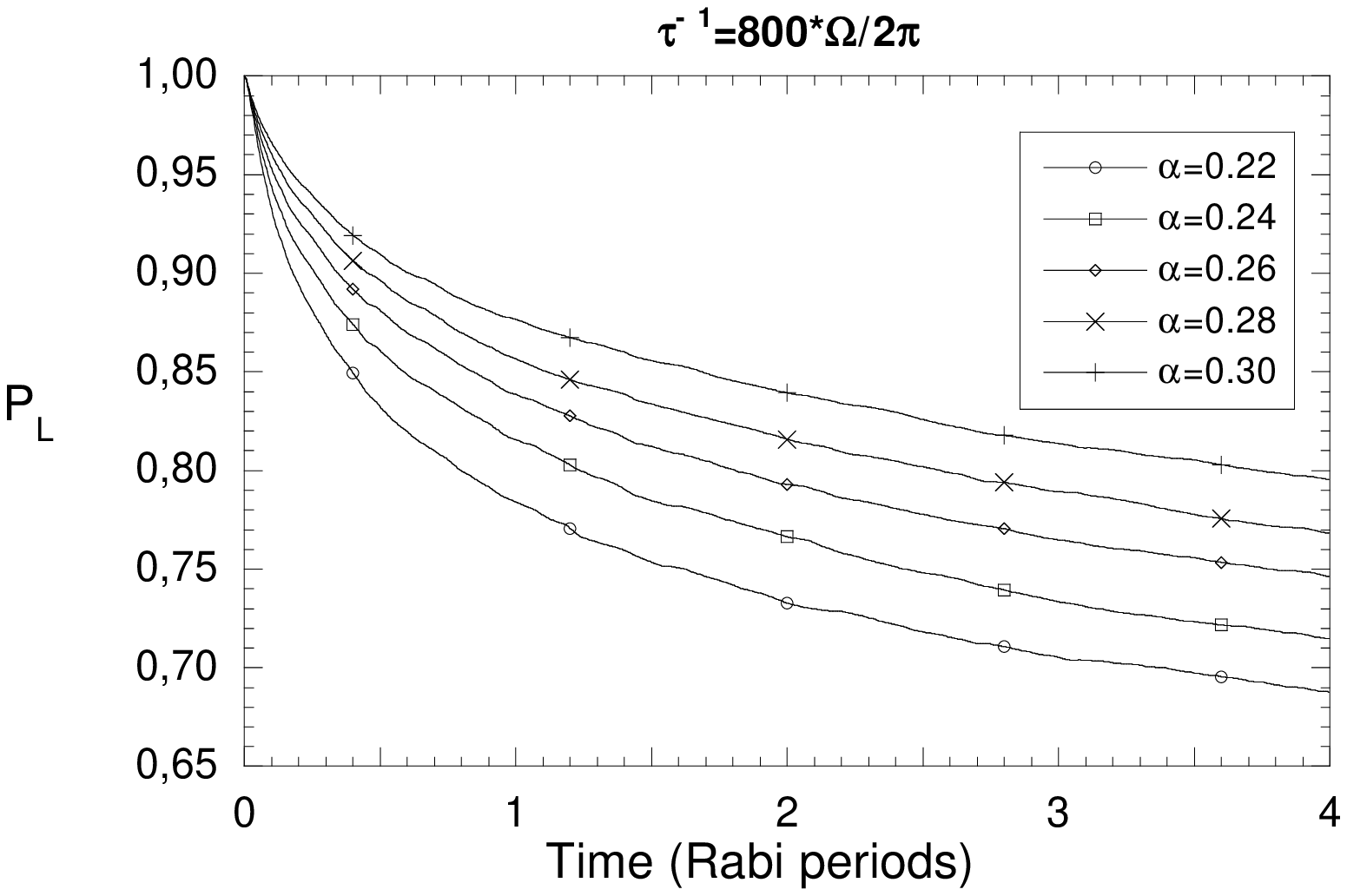}
\caption{\label{fig:fig2} Temporal evolution of $P_L$ as a
function of $\alpha=\alpha_L=\alpha_R$. For all calculations we set
$\tau^{-1}=800 \Omega/2\pi$. }
\end{figure}

A similar Zeno effect is evident when the parameter $\alpha$ is
varied, while keeping the collision frequency $\tau^{-1}$ constant,
as displayed in \ref{fig:fig2} (once again, we only display
the numerical results, for the analytical ones cannot be
distinguished). Unlike in the preceding case, where the Zeno effect
was due to increasing collision \emph{frequency}, now it is due to
increasing collision \emph{effectiveness}: a larger $\alpha$ entails
more dephasing and decoherence and, in a loose sense, a better
"measurement" of the quantum state. The parameter $x=\alpha^2
T/\tau$ ranges between 39 and 72 ($\gg 1$) and one observes again
the presence of a (parameter-independent) short-time region.

\begin{figure}
\includegraphics{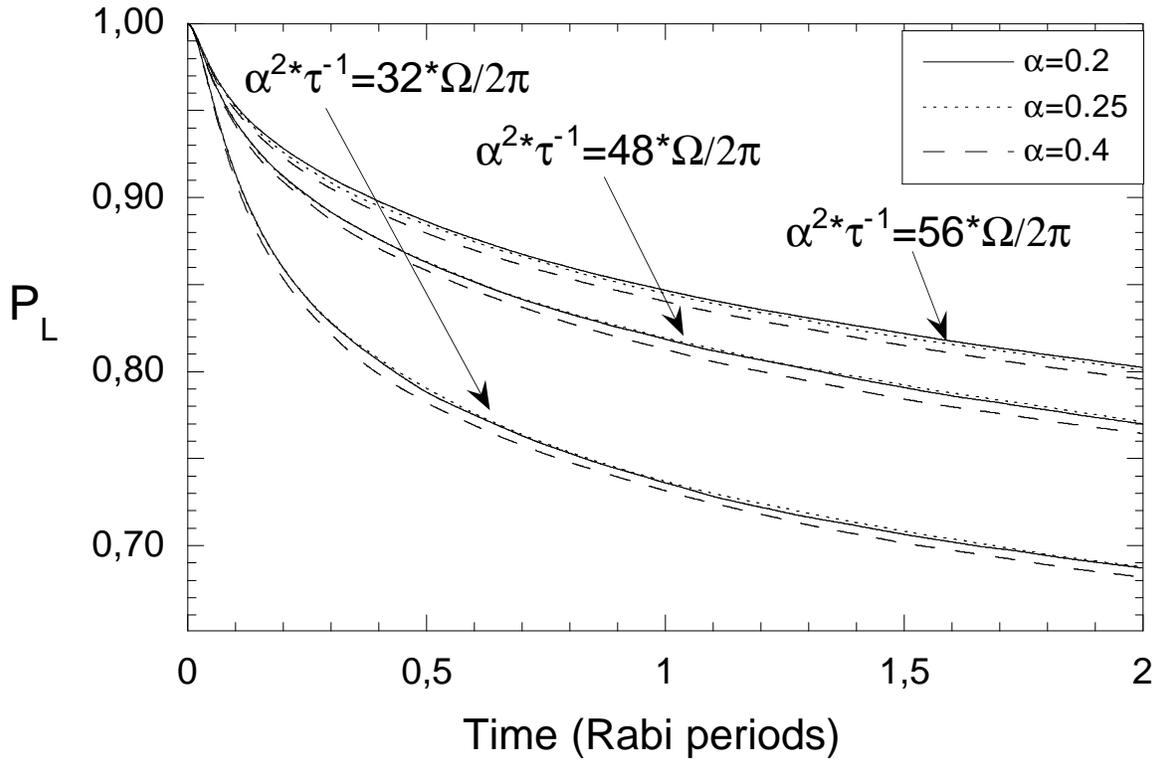}
\caption{\label{fig:fig4} Test of the scaling law (\ref{eq:beta}).
Temporal evolution of $P_L$ for different values of $x=\alpha^2
T/\tau$: three simulations were done with
$\alpha_R=\alpha_L=\alpha=0.2,\ 0.25,\ 0.4$, respectively. $x$
ranges between 32 to 56.}
\end{figure}

As the analysis of Secs.
\ref{sec-master}-\ref{sec-zenoeffcoupl} shows,
the dynamics of the system should be ruled by the scaling parameter
\beq
\label{eq:beta}
x=D T_R=\frac{\alpha^2 T_R}{\tau}.
\eeq
\ref{fig:fig4} shows how this scaling law is supported by the
results of the numerical simulation. The plot shows three sets of
curves corresponding to three different values of $x$. In each set,
the values of $\alpha$ and $\tau$ were varied as indicated. Some
deviations from the scaling law (of order 5\%) can be observed and
are to be ascribed to the influence of the terms neglected in
deriving Eq.\ (\ref{eq:master}). Incidentally, notice again the
short-time quadratic behavior.

\section{Concluding remarks}
\label{sec-concrem}\andy{sec-concrem}

We have studied a Zeno effect in a multilevel molecule made up of 
40+40 levels, one of which (the ground
state of the left subspace) is initially populated, and the
evolution towards the right subspace is slowed down both because the
collisions remove population density "upwards" from the left ground
state (a classically intuitive process) and because they "dephase"
(or analogously, make energetically less favorable) the transitions
towards the right subspace. The latter process is classically less
intuitive, but is readily understood if one thinks in terms of
quantum transition amplitudes (or of the Fermi "golden" rule for a
\emph{bona fide} unstable system).

It is worth stressing that the general ideas and techniques
introduced in this article are valid for any multilevel molecule and
any possible level distribution: we focused on the case
(\ref{eq:energyrot}) only for concreteness. Those situations in
which (\ref{eq:reduced}) is not valid are very particular cases and
their analysis, although of interest, goes beyond the scope of this
article.

On the other hand, it is also necessary to emphasize that we
neglected temperature effects and rapid structural rearrangement
phenomena leading to a Boltzmann distribution of the level
populations. This is a conceptually interesting problem, that involves 
delicate issues: a sensible
estimate of the timescales involved in these thermalization
processes is a challenging problem that requires further
investigation.

We conclude by noticing that the Hamiltonian
(\ref{eq:Htot})-(\ref{Hdetailscoll1}) is also relevant for the study
of quantum chaos and Anderson localization \cite{chaos}. The
analysis of Poissonianly distributed "kicks" (\ref{eq:avtau1})
would introduce a novel element of discussion in such a context.

\begin{acknowledgement}

P.F.\ and S.P.\ acknowledge the financial support of the European
Union through the Integrated Project EuroSQIP.
D.B., S.L.\ and P.M.\ were partially supported by Ministero
dell'Istruzione, dell'Universit\`{a} e della Ricerca (Contract 2001031223\_009).

\end{acknowledgement}

\section{Appendix}
\appendix
\label{sec-appA}

It is interesting to look explicitly at the derivation of Eq.\
(\ref{eq:master}) from Eq.\ (\ref{eq:genme}). The physical mechanism
at work is the effective decoupling between the fast and the slow
modes in (\ref{eq:genme}). Let us start from the equation for
$\rho_{1_L2_L}$, that explicitly reads [here
$\omega_{2_L1_L}\equiv\left(E_{2_L}-E_{1_L}\right)/\hbar$]
\andy{totdyn}
\barr
\frac{d\rho_{1_L2_L}}{dt}&=&i\omega_{2_L
1_L}\rho_{1_L2_L}-i\Omega\rho_{1_R2_L}-i\frac{\alpha_L}{\tau}\left(\rho_{2_L
2_L}-\rho_{1_L1_L}-\rho_{1_L3_L}\right)\nonumber\\
&&-\frac{\alpha^2_L}{2\tau}(\rho_{1_L2_L}+\rho_{3_L2_L}
-2\rho_{2_L1_L}-2\rho_{2_L3_L}+2\rho_{1_L2_L}+\rho_{1_L4_L}).
\label{eq:totdyn}
\earr
When condition (\ref{eq:reduced}) is satisfied, the first term in
the right-hand side dominates over the others and one obtains
\andy{fastdyn}
\beq
\frac{d\rho_{1_L2_L}}{dt}\simeq i\omega_{2_L1_L}\rho_{1_L2_L},
\label{eq:fastdyn}
\eeq
which yields a very fast dynamics for the term $\rho_{1_L2_L}$:
\beq
\rho_{1_L2_L}(t)=\rho_{1_L2_L}(0)\exp({i\omega_{2_L1_L}t}).
\label{eq:fastdyn3}
\eeq
The equations for the other off-diagonal components of $\rho$ are
similar. These equations yield very rapidly oscillating solutions.

On the other hand, the dynamics of the populations
$\rho_{1_s1_{s}}$, with $s=L,R$, and of the coherent terms
$\rho_{1_L 1_R}$ is governed by the equations
\barr
\frac{d\rho_{1_L 1_L}}{dt}&=& -i\Omega (\rho_{1_R 1_L}-\rho_{1_L
1_R}) + \frac{\alpha_L^2}{\tau}(\rho_{2_L 2_L}-\rho_{1_L 1_L}) -i
\frac{\alpha_L}{\tau} (\rho_{2_L 1_L}-\rho_{1_L 2_L})
-\frac{\alpha^2_L}{2\tau}(\rho_{3_L 1_L}+ \rho_{1_L 3_L}),
\nonumber\\
\frac{d\rho_{1_R 1_R}}{dt}&=&  i\Omega (\rho_{1_R 1_L}-\rho_{1_L
1_R}) + \frac{\alpha_R^2}{\tau}(\rho_{2_R 2_R}-\rho_{1_R 1_R}) -i
\frac{\alpha_R}{\tau} (\rho_{2_R 1_R}-\rho_{1_R 2_R})
-\frac{\alpha^2_R}{2\tau}(\rho_{3_R 1_R}+ \rho_{1_R 3_R}),
\nonumber \\
\frac{d\rho_{1_L 1_R}}{dt}&=&  i\Omega (\rho_{1_L 1_L}-\rho_{1_R
1_R}) - \frac{\alpha_L^2}{2\tau}\rho_{1_L 1_R}-
\frac{\alpha_R^2}{2\tau}\rho_{1_R 1_L} -i \frac{\alpha_L}{\tau}
\rho_{2_L 1_R} -i \frac{\alpha_R}{\tau} \rho_{1_L 2_R}
+\frac{\alpha_L\alpha_R}{\tau} \rho_{2_L 2_R}
\nonumber \\
& &
-\frac{\alpha^2_L}{2\tau}\rho_{3_L 1_R}
-\frac{\alpha^2_R}{2\tau}\rho_{1_L 3_R} .
\earr
It is apparent that no "diagonal" fast frequency $\omega$ is
present and these matrix elements evolve over timescales $\tau$ and
$\Omega^{-1}$ which are much larger than $\omega^{-1}$. Therefore
the contribution of all the off-diagonal fast terms of the type
(\ref{eq:fastdyn3}) is averaged to zero over the long timescales
$\tau$ and $\Omega^{-1}$, the dynamics of the slow and fast terms
completely decouples and we get
\barr
\frac{d\rho_{1_L 1_L}}{dt}&\simeq& -i\Omega (\rho_{1_R
1_L}-\rho_{1_L 1_R}) + \frac{\alpha_L^2}{\tau}(\rho_{2_L
2_L}-\rho_{1_L 1_L}),
\nonumber\\
\frac{d\rho_{1_R 1_R}}{dt}&\simeq& i\Omega (\rho_{1_R
1_L}-\rho_{1_L 1_R}) + \frac{\alpha_R^2}{\tau}(\rho_{2_R
2_R}-\rho_{1_R 1_R}),
\nonumber\\
\frac{d\rho_{1_L 1_R}}{dt}&\simeq&  i\Omega (\rho_{1_L
1_L}-\rho_{1_R 1_R}) - \frac{\alpha_L^2}{2\tau}\rho_{1_L 1_R}-
\frac{\alpha_R^2}{2\tau}\rho_{1_R 1_L}.
\nonumber\\
\earr

Analogously, the evolution equations of the populations
$p_{m_s}=\rho_{m_sm_s}$ read ($m_s\neq 1_{L,R}$)
\barr
\frac{d\rho_{m_sm_s}}{dt}&=&
\frac{\alpha^2_s}{\tau}\left(\rho_{m_s-1,m_s-1}-2\rho_{m_sm_s}+
\rho_{m_s+1,m_s+1}\right)
\nonumber \\
& & 
-i\frac{\alpha_s}{\tau}\left(\rho_{m_s+1,m_s}
-\rho_{m_s,m_s+1}+\rho_{m_s-1, m_s}-\rho_{m_s,m_s-1} \right)
\nonumber\\
& & -\frac{\alpha^2_s}{2\tau}\left(\rho_{m_s-2,m_s}+
\rho_{m_s,m_s-2}+\rho_{m_s+2,m_s}+ \rho_{m_s,m_s+2} \right),
\earr
and by the same argument reduce to
\barr
\frac{d\rho_{m_sm_s}}{dt}\simeq
\frac{\alpha^2_s}{\tau}\left(\rho_{m_s-1,m_s-1}-2\rho_{m_sm_s}+
\rho_{m_s+1,m_s+1}\right),
\nonumber\\
\earr
which are in the form (\ref{eq:reducedme})-(\ref{eq:master}). Notice
the absence of fast and oscillating terms.


\bibliography{achemso}

\end{document}